\documentclass[11pt,a4paper]{article}
\usepackage[T1]{fontenc}
\usepackage[utf8]{inputenc}
\usepackage[english]{babel}
\usepackage{authblk}
\usepackage{amsthm}
\usepackage{color}
\usepackage[colorlinks=true,urlcolor=blue,linkcolor=blue]{hyperref}
\usepackage{graphicx}
\usepackage[bottom]{footmisc}
\usepackage{fullpage}
\usepackage{calc}
\usepackage{amsmath,amssymb}
\usepackage{amsfonts}
\usepackage{mathrsfs}
\usepackage[nottoc]{tocbibind}
\usepackage{graphicx}
\graphicspath{ {images/} }
\usepackage{fancyhdr}
\usepackage{tabularx}
\usepackage{array}
\usepackage{lscape}
\usepackage{array,multirow,makecell}
\setcellgapes{1pt}
\makegapedcells
\usepackage[table]{xcolor}
\usepackage{float}
\usepackage{caption}
\usepackage{subcaption}
\usepackage{pifont}
\usepackage[ruled,vlined,linesnumbered,noresetcount]{algorithm2e}
\usepackage{booktabs}
\usepackage{titlesec}
\usepackage[section]{placeins}
\usepackage{bbm}
\hypersetup{citecolor=blue}
\newcolumntype{R}[1]{>{\raggedleft\arraybackslash }b{#1}}
\newcolumntype{L}[1]{>{\raggedright\arraybackslash }b{#1}}
\newcolumntype{C}[1]{>{\centering\arraybackslash }b{#1}}

\newtheorem{defi}{Definition}

\newtheorem{rem}{Remark}

\makeatletter
\renewcommand\paragraph{\@startsection{paragraph}{4}{\z@}
  {-3.25ex \@plus -1ex \@minus -0.2ex}
  {2.25ex \@plus .25ex}
  {\normalfont\normalsize\bfseries}}
\renewcommand\subparagraph{\@startsection{subparagraph}{5}{\z@}
  {-3.25ex \@plus -1ex \@minus -0.2ex}
  {2.25ex \@plus .25ex}
  {\normalfont\normalsize\bfseries}}
\def\toclevel@paragraph{4}
\def\toclevel@paragraph{5}
\def\l@paragraph{\@dottedtocline{4}{7em}{4em}}
\def\l@subparagraph{\@dottedtocline{5}{7em}{4em}}
\makeatother

\title{Market Impact: A Systematic Study of the High Frequency Options Market \\
\small \textit{(published in Quantitative Finance, Volume 21, 2021 - Issue 1)}}
\author[*,***]{Emilio Said}
\author[*]{Ahmed Bel Hadj Ayed}
\author[*]{Damien Thillou}
\author[*]{Jean-Jacques Rabeyrin}
\author[**,***]{Frédéric Abergel}

\affil[*]{Quantitative Research, Global Markets, BNP Paribas, Paris, France}
\affil[**]{Quantitative Research Group, BNP Paribas Asset Management, Paris, France}
\affil[***]{Chaire de Finance Quantitative, Laboratoire MICS, CentraleSupélec, Université Paris-Saclay, Gif-Sur-Yvette, France}

\begin{document}

\maketitle

\begin{abstract}
This paper deals with a fundamental subject that has seldom been addressed in recent years, that of market impact in the options market. Our analysis is based on a proprietary database of metaorders - large orders that are split into smaller pieces before being sent to the market - on one of the main Asian markets. In line with our previous work on the equity market \cite{said2018}, we propose an algorithmic approach to identify metaorders, based on some implied volatility parameters, the \textit{at the money forward volatility} and \textit{at the money forward skew}. In both cases, we obtain results similar to the now well understood equity market: \textit{Square-Root Law}, \textit{Fair Pricing Condition} and \textit{Market Impact Dynamics}.
\end{abstract}


\section{Introduction}
\label{introduction}

In recent years market impact has become a topic of interest for most market participants. The advent of algorithmic trading has significantly increased the traded volumes and the number of transactions. The whole point of electronic markets is to directly match participants that are willing to sell an asset with participants that are willing to buy it. This is mainly done via two types of orders: market orders and limit orders. Market orders are sent by participants that are willing to either buy or sell the asset immediately. Limit orders, however, do not share this urgency: these orders show the interest of the participant to buy or sell the asset at a pre-assigned price. Market orders are generally not used by institutional investors because of the lack of control they imply. On the contrary, limit orders, whether they are \emph{aggressive} - crossing the spread - or \emph{passive}, form the vast majority of orders actually sent to the market during the execution of an algorithmic trading strategy, for example a market making strategy or an optimal execution strategy.
Most strategies referred to as algorithmic trading fall into the cost-reduction category. The basic idea is to break down a large order -- a \textbf{metaorder} -- into small orders and send them to the market over time. The reasons for the incremental execution of metaorders are originally due to what we have called the \textit{liquidity paradox}: The volume of buy or sell limit orders typically available in the order book
at a given instant of time is quite small and represents only the order of $1\%$ of the traded daily volume, i.e. $10^{-4}-10^{-5}$ of the market cap for stocks \cite{bouchaud2009markets}. Hence the fact that the outstanding liquidity is so small has an immediate consequence: trades must be fragmented.

Recently, market impact questions related to options market have appeared in option pricing issues. However, the constraints of non-arbitrage often impose some limitations such as linear market impact \cite{bouchard2017hedging} \cite{loeper2018option}. In these papers the authors examine how linear market impact on the spot market can affect option prices. The point of view adopted here is slightly different and seeks to highlight how dealing with options can affect directly volatility and therefore option prices.

Many studies have been conducted to understand the influences of metaorders on the price formation process. Most of them are concerned with the equity markets, see e.g. \cite{almgren2005direct}, \cite{moro2009market}, \cite{toth2011anomalous}, \cite{bershova2013non}, \cite{mastromatteo2014agent}, \cite{bacry2015market}, \cite{gomes2015market}, \cite{brokmann2015slow} and \cite{said2018}. All these studies have shown a common behavior during the execution of a metaorder, namely, a concave and temporary impact followed by a convex and decreasing relaxation. More recently, \cite{donier2015million} have observed similar effects in the bitcoin market, and a short note \cite{toth2016square} indicated that the \textit{Square root law} seems to be hold for the options market.

This paper is intended as an analysis of market impact in the options market. To the best of our knowledge, this is the first detailed, in-depth academic study of this phenomenon.

The paper is organized as follow: Section \ref{definitions, algorithm and market impact measures} recalls our algorithmic definition of an \textit{equity metaorder}, introduces that of an \textit{options meatorder} and presents some market impact measures. Section \ref{data} introduces the data set of the study and presents our approach for \textit{options} metaorders. Sections \ref{market impact dynamics}, \ref{square root law} and \ref{fair pricing} present our findings and empirical results: They confirm that the market impact law observed in the equity markets also hold true in the options markets. Section \ref{conclusion} is a discussion of our results and their implications.

\section{Definitions, Algorithm and Market Impact measures}
\label{definitions, algorithm and market impact measures}

\subsection{Basic Definitions}
\label{basic definitions}

Some basic concepts, and the algorithmic definition of an options metaorder, are introduced here.

\begin{defi}
A \textbf{limit order} is an order that sets the maximum or minimum price at which an agent is willing to buy or sell a given quantity of a particular stock.
\end{defi}

\begin{defi}
An \textbf{aggressive limit order} is one that instantaneously removes liquidity from the order book by triggering a transaction. An aggressive order crosses the Bid–Ask spread. In other words an aggressive buy order will be placed on the ask, and an aggressive sell order will be placed on the bid.
\end{defi}

A limit order that is not aggressive is termed \textbf{passive}. Passive orders sit in the order book until they are executed or cancelled.

Loosely speaking, a \textbf{metaorder} is a large trading order that is split into small pieces and executed incrementally. In order to perform rigorous statistical analyses, a more specific and precise definition of a metaorder is required, and given in Definition \ref{defMetaorder} below:

\begin{defi}
\label{defMetaorder}
A \textbf{metaorder} is a series of orders sequentially executed during the same day and having those same attributes:
\begin{itemize}
    \item \textcolor{blue}{agent} i.e. a participant on the market (an algorithm, a trader...);
    \item \textcolor{blue}{product id} i.e. a financial instrument (a share, an option...);
    \item \textcolor{blue}{direction} (buy or sell);
\end{itemize}
\end{defi}

Clearly, Definition \ref{defMetaorder} must be adapted to fit the options market.

Options are a bit more complex than equities. Traders buy and sell volatility and deal directly with the implied volatility surface, and therefore, with their implicit volatility parameters. As such, an options metaorder can naturally be defined as a sequence of transactions that generate some specific deformations of the volatility surface.

\begin{defi}
\label{defMetaorderOption}
An \textbf{options metaorder} with respect to an implied volatility parameter $\theta$ is a series of orders sequentially executed during the same day and having those same attributes:
\begin{itemize}
    \item \textcolor{blue}{agent} i.e. a participant on the market (an algorithm, a trader...);
    \item \textcolor{blue}{underlying product id} i.e. the underlying financial instrument;
    \item \textcolor{blue}{direction} regarding the sign of $\mathcal{S}^{\theta} := Q \times \displaystyle\frac{\partial \mathcal{O}}{\partial \theta}$ where $Q$ is the algebraic quantity (positive for a buy order and negative for a sell order) and $\mathcal{O}$ the price of the option traded;
\end{itemize}
\end{defi}

Note that in Definition \ref{defMetaorderOption}, the \textit{product id} condition introduced in Definition \ref{defMetaorder} is dropped. As a matter of fact, trading an option with a given strike $K$ and maturity $T$ also affects those with nearby strikes and maturities, so that trades on options with different strikes and maturity can very well belong to the same metaorder.

This approach is in line with what is presented in \cite{said2018}, and leads to a systematic study of options market impact.

In what follows we will use the term $\theta-$metaorder to refer to an options metaorder with respect to the implied volatility parameter $\theta$ as defined in Definition \ref{defMetaorderOption}.

\subsection{Market Impact definitions}
\label{market impact definitions}

The framework is similar to that introduced in \cite{said2018}. Let $\Omega^{\theta}$ be the set of $\theta$-metaorders under scrutiny, that is, $\theta-$metaorders that are fully executed during a single market session, and pick $\omega \in \Omega^{\theta}$ executed on (possibly) several options with the same underlying and during a given day $d$. Its execution starts at some time $t_0(\omega)$ and ends the same day at time $t_0(\omega) + T(\omega)$. Thus $T(\omega)$ represents the duration of the metaorder. Denote by N($\omega$) the number of orders that have been executed during the life cycle of the metaorder $\omega$: $N(\omega)$ is the length of $\omega$. Let $t_0(\omega), t_1(\omega), ..., t_{N(\omega)-1}(\omega)$ be the transaction times of the metaorder $\omega$, we define $ \mathcal{V}^{\theta}(\omega) := \displaystyle\sum_{i=0}^{N(\omega) - 1}{\mathcal{S}_{t_i(\omega)}^{\theta}} $ as the sensitivity of the metaorder $\omega$ regarding to the parameter $\theta$. Let $V^{\theta} := \displaystyle\sum_{t \in \mathcal{T}_{market}}{\left|\mathcal{S}_t^{\theta}\right|} $ be the sensitivity traded the same day $d$ on all the options of the universe, which means all the options traded by the algorithms summed over all market transactions $\mathcal{T}_{market}$ occurred in the day $d$. Hence $V^{\theta}$ can be viewed as the \textit{absolute sensitivity} traded by the market regarding to the parameter $\theta$. Note that this quantity depends only on the universe the day $d$. Therefore all the $\theta-$metaorders executed the same day will share the same \textit{absolute sensitivity}. Hence we define $\displaystyle\frac{\left|\mathcal{V}^{\theta}(\omega)\right|}{V^{\theta}}$ as the $\theta-$daily participation rate. The sign of $\omega$ will be noted $\epsilon(\omega)$ (i.e. $\epsilon = 1$ for a positive sensitivity options metaorder and $\epsilon = -1$ for a sensitive negative one), the sign of $\epsilon(\omega)$ is also the sign of $\mathcal{S}_{t_0(\omega)}^{\theta}, \mathcal{S}_{t_1(\omega)}^{\theta}, ..., \mathcal{S}_{t_{N(\omega) - 1}(\omega)}^{\theta}$ which is invariant during the life of $\omega$. Clearly, most of the quantities introduced in this section depend on $\omega$. For the sake of simplicity, we chose to omit this dependence whenever there is no ambiguity and will often write $T$, $N$, $\mathcal{V}^{\theta}$, $\epsilon$ instead of $T(\omega)$, $N(\omega)$, $\mathcal{V}^{\theta}(\omega)$, $\epsilon(\omega)$.

The \textbf{market impact curve} of a metaorder $\omega$ quantifies the magnitude of the relative $\theta-$variation between the starting time of the metaorder $t_0$ and the current time $t > t_0$, $\theta$ being a parameter of the implied volatility model. Let $\mathcal{I}_{t}(\omega)$ be a proxy for the realized $\theta-$parameter variation between time $t_0$ and time $t_0 + t$. We use the \textit{variation proxy} defined by
\begin{equation}
\label{variation proxy}
\mathcal{I}_{t} = \theta_t - \theta_{t_0},
\end{equation}
This estimation relies on the assumption that the \textit{exogenous market moves} $W_t$ will cancel out once averaged, i.e. as a random variable, $W_t$ should have finite variance and basically satisfy $\mathbb{E}(\epsilon(\omega)W_{t}(\omega)) = 0 $.

One can thus write
\begin{equation}
\epsilon(\omega)\mathcal{I}_{t}(\omega) = \eta_{t}(\omega) + \epsilon(\omega)W_{t}(\omega),
\end{equation}
where $\eta_{t}(\omega)$ represents the market impact curve and $W_{t}(\omega)$, the exogenous variation corresponding to the relative move that would have occurred if the metaorder had not been sent to the market.

\begin{center}
\begin{tabular}{l c c}
\toprule[0.15 em]
 & \multicolumn{1}{c}{Equity Metaorder} & \multicolumn{1}{c}{Options Metaorder}\\
\midrule[0.1 em]
\centering Object of study & \centering stock market & \centering options market \tabularnewline
\centering Quantity of interest & \centering stock price $P$ & \centering implied volatility parameters $\theta$ \tabularnewline
\centering Effective size & quantity \centering $Q$ & \centering sensitivity $\mathcal{S}^{\theta}$ \tabularnewline
\centering Effective variation & \centering $P_t - P_{t_0}$ & \centering $\theta_t - \theta_{t_0}$ \tabularnewline
\centering Effective Sign $\epsilon$ & \centering $\mathbbm{1}_{\{Q > 0\}} - \mathbbm{1}_{\{Q < 0\}}$ & \centering $\mathbbm{1}_{\{\mathcal{S}^{\theta} > 0\}} - \mathbbm{1}_{\{\mathcal{S}^{\theta} < 0\}}$ \tabularnewline
\centering Market impact proxy & \centering $\epsilon \times (P_t - P_{t_0})$ & \centering $\epsilon \times (\theta_t - \theta_{t_0})$ \tabularnewline
\bottomrule[0.15 em]
\end{tabular}
\captionof{table}{\textit{Comparison of the quantities of interest between equity and options metaorders}}
\label{tab empirical studies}
\end{center}

\subsection{Implied volatility model}
\label{implied volatility model}

In the previous section we have introduced the following proxy
\begin{equation}
\epsilon \times (\theta_t - \theta_{t_0})
\end{equation}
to measure the market impact of $\theta$ which is a parameter of the proprietary implied volatility model used to conduct this study. Although the approach used is model dependent, we think that for the two parameters presented here (\textit{at the money forward} volatility and skew) the results must be the same for any implied volatility model fitting the market. This is also one of the reasons why we limited ourselves to study the market impact on \textit{the at the money forward} volatility and skew in our article as every reasonnable implied volatility model must have at least those two variables. We have respectively sketched on Figures \ref{simple vol atmf strikes} and \ref{simple ATMF skew strikes} the expected market impact variations of the \textit{at the money forward} volatility and skew. For sake of clarity let us recall that the terminology \textit{at the money forward} refers to quantities function of the strike evaluated at the forward.

\section{Data}
\label{data}

\subsection{Data description}
\label{data description}

The data set we use for this analysis contains trade orders executed by the BNP Paribas options trading desk for the 2-year period from June 2016 through June 2018 on the KOSPI 200 options. The KOSPI 200 Index is a capitalization-weighted index of 200 Korean stocks which make up 93\% of the total market value of the Korea Stock Exchange. In order to perform rigorous statistical analyses we need to be able to calibrate the parameters of the implied volatility model as often as possible. This is necessary to observe the variations of the parameters at a frequency similar to that of the transactions. To make this possible with sufficient accuracy, we have only considered executions trading mostly short maturities options.

Because of the high frequency in the execution of the orders, \textbf{we only consider options metaorders with at least 5 completed transactions}. This underlines the fact that we want keep only metaorders that reasonably act as liquidity takers and could impact significantly the market. Indeed, while \textbf{the equity metaorders studied in \cite{said2018} had an average time life of several hours, the options metaorders presented here last a few tenths of seconds}.

\subsection{Data representation}
\label{Data representation}

For sake of confidentiality numerical values of market impact, daily participation rates and durations are given in percentage of reference values -- related to the market and BNP Paribas options desk -- that cannot be disclosed. Nevertheless we will give when it is possible the value of the ratio of these reference values in order to make feasible comparisons between the \textit{at the money forward} volatility and skew metaorders. For instance the reference value of impact is the same in Figures \ref{vol atmf dynamics N_Vol 5}, \ref{vol atmf dynamics N_Vol 10} and \ref{vol atmf dynamics N_Vol 15} for the volatility metaorders, so it is possible to make comparisons between these graphs. The same applies for skew metaorders in Figures \ref{ATMF skew dynamics N_Smile 5}, \ref{ATMF skew dynamics N_Smile 10} and \ref{ATMF skew dynamics N_Smile 15}.

In what follows we will use the following convention: All graphs relating to volatility (resp. skew) metaorders will be drawn in blue (resp. green).

\subsection{Options Market Impact - The Liquidity Taker Mode (Aggressive Orders)}
\label{the liquidity taker mode (aggressive orders)}

Let us now focus on the market impact generated by a series of aggressive executions on the options market. More specifically, we consider the two kinds of metaorders: namely the \textit{at the money forward volatility} and the \textit{at the money forward skew} metaorders. We recall that aggressive limit orders are limit orders that cross the spread in order to trigger an immediate transaction.

\begin{figure}
\centering
\includegraphics[scale=0.60]{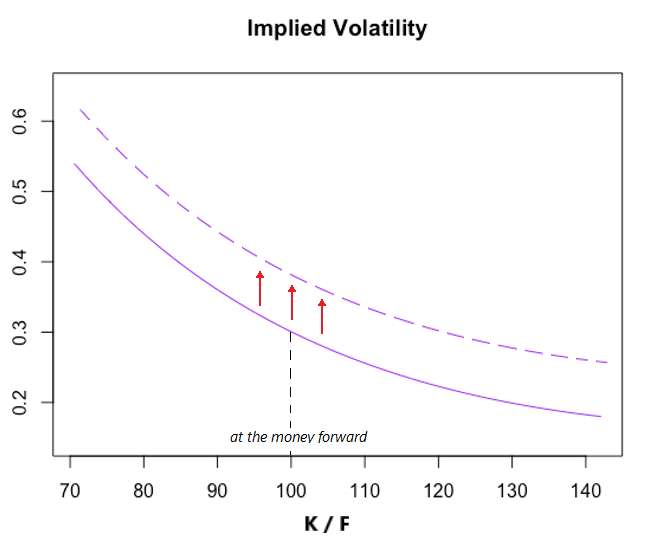}
\captionof{figure}{Market impact on the implied volatility surface for a given maturity $T$ under the effect of buying options with strikes near the money forward. One can notice how this could increase the \textit{at the money forward volatility} parameter as emphasized by the red arrows.}
\label{simple vol atmf strikes}
\end{figure}

\begin{figure}
\centering
\includegraphics[scale=0.60]{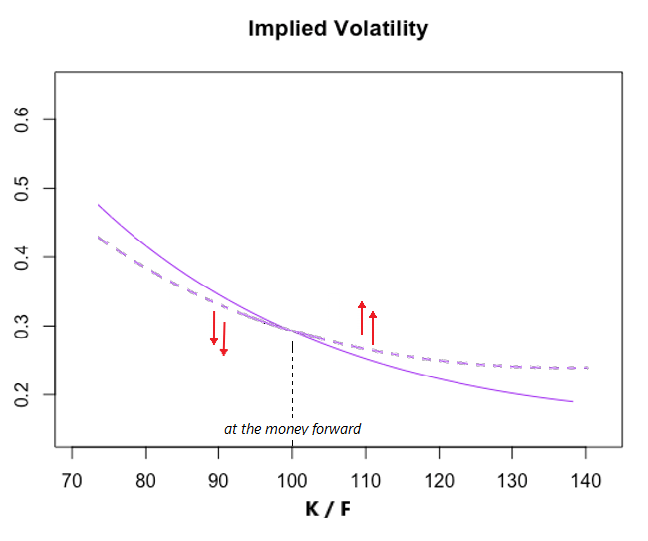}
\captionof{figure}{Market impact on the implied volatility surface for a given maturity $T$ under the effect of buying options with strikes at the right of the money forward and sell options with strikes at the left of the money forward at the same time. One can see how this could increase the \textit{at the money forward skew} parameter as emphasized by the red arrows.}
\label{simple ATMF skew strikes}
\end{figure}

It is important to understand the fundamental difference between these two types of metaorders. For example, if someone wants to buy \textit{at the money forward volatility} (Fig. \ref{simple vol atmf strikes}), the simplest strategy is to buy options with strikes close to the money forward. Whereas if one wants to buy \textit{at the money forward skew}, a simple way to do so consists in buying out of the money options while at the same time, selling in the money options (Fig. \ref{simple ATMF skew strikes}).

\begin{figure}[H]
\centering
\includegraphics[scale=0.80]{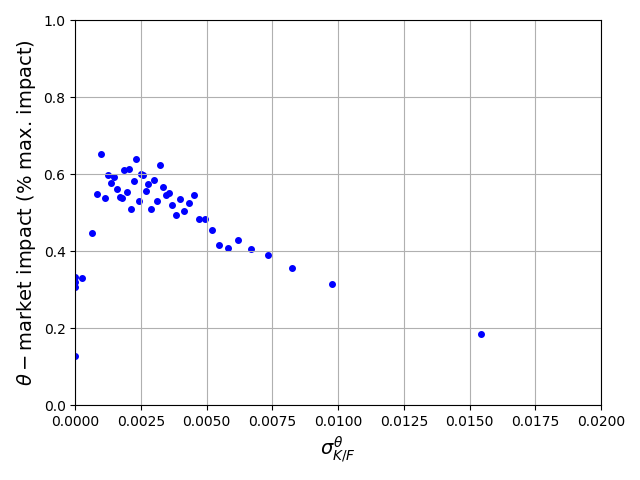}
\captionof{figure}{Market impact as a function of $\sigma_{K/F}$ in the case of the \textit{at the money forward volatility} metaorders ($\theta \equiv$ ATMF volatility)}
\label{strikes ATMF vol}
\end{figure}

\begin{figure}[H]
\centering
\includegraphics[scale=0.80]{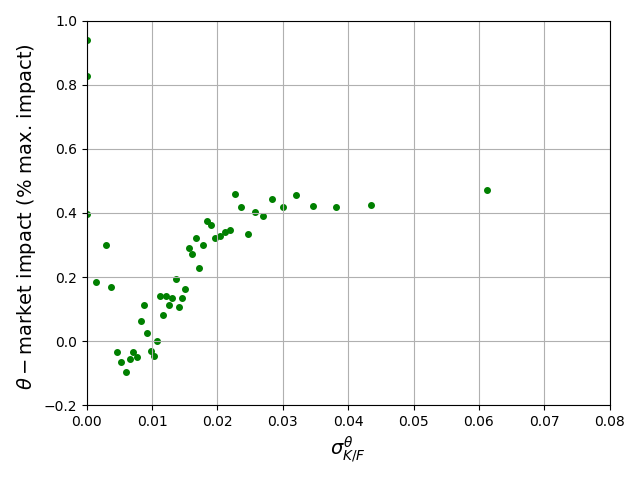}
\captionof{figure}{Market impact as a function of $\sigma_{K/F}$ in the case of the \textit{at the money forward skew} metaorders ($\theta \equiv$ ATMF skew)}
\label{strikes ATMF skew}
\end{figure}

A simple way to check that the previous intuitions presented in Figures \ref{simple vol atmf strikes} and \ref{simple ATMF skew strikes} are not totally irrelevant is to study for the two kinds of metaorders the relation of the $\theta-$market impact curve as a function of the (unbiased) standard deviation (weighted by the $\theta$-sensitivity) of the strikes of the child orders pertaining for each metaorder denoted by $\sigma_{K/F}$ (Fig. \ref{strikes ATMF vol} and \ref{strikes ATMF skew}). Hence for each $\theta-$metaorder $\omega$ the expression of $\sigma_{K/F}^{\theta}(\omega)$ is given by
\begin{equation}
  \sigma_{K/F}^{\theta}(\omega) = \sqrt{\frac{\displaystyle\sum_{i=0}^{N(\omega)-1}{\mathcal{S}_{t_i(\omega)}^{\theta}\big(K_i / F_{t_i(\omega)} - \mu_{K/F}^{\theta}(\omega)\big)^2}}{\mathcal{V}^{\theta}(\omega) - 1}},
\end{equation}
where
\begin{equation}
  \mu_{K/F}^{\theta}(\omega) = \frac{\displaystyle\sum_{i=0}^{N(\omega)-1}{\mathcal{S}_{t_i(\omega)} K_i / F_{t_i(\omega)}}}{\mathcal{V}^{\theta}(\omega)}
\end{equation}
is the weighted mean with $K_i$ the strikes of the corresponding child orders and $F_{t_i}$ the forward prices at the instants $t_i$. Figures \ref{strikes ATMF vol} and \ref{strikes ATMF skew} are obtained by following the methodology presented in \cite{said2018} and briefly recalled in Sec. \ref{market impact curves}.

A careful scrutiny of Figures \ref{strikes ATMF vol} and \ref{strikes ATMF skew} shows that in the case of the \textit{at the money forward volatility metaorders}, the less diversified the strikes are, the more important the market impact. It is the opposite in the case of the \textit{at the money forward skew metaorders}: the metaorders with the higher values of $\sigma_{K/F}$ match those which present the most important market impacts. Besides as expected, one can note that the average values of $\sigma_{K/F}^{\theta}$ are higher for the \textit{at the money skew metaorders}. Those observations are in line with the intuitive predictions presented in Figures \ref{simple vol atmf strikes} and \ref{simple ATMF skew strikes}. The first points appearing on the Figures \ref{simple vol atmf strikes} and \ref{simple ATMF skew strikes} actually correspond to the shortest metaorders in length and duration, so they are just noisy points.

\subsubsection{The ATMF volatility metaorders}
\label{the ATMF volatiliy metaorders}

In this section we focus on metaorders impacting the \textit{at the money forward volatility} parameter of the implied volatility model.

\paragraph{Data}
\label{data ATMF volatility metaorders}

\begin{itemize}
    \item Study period : \textbf{1st July 2016 -- 30th June 2018}
    \item Order types : \textbf{Aggressive Limit Orders}
    \item Parameter : $\theta \equiv$ \textbf{at the money forward volatility}
    \item Filters : \textbf{metaorders $\omega \in \Omega^{\theta}$}
    \item Number of orders : \textbf{1,026,197}
    \item Number of metaorders : \textbf{149,441}
\end{itemize}

\paragraph{Duration distribution}
\label{duration distribution ATMF volatility metaorders}

\begin{figure}[H]
\centering
\includegraphics[scale=0.80]{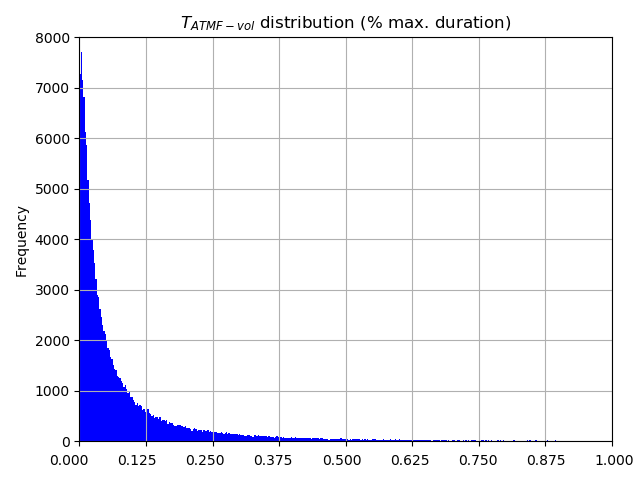}
\captionof{figure}{Duration distribution of the \textit{at the money forward volatility} metaorders}
\label{vol atmf T distribution}
\end{figure}

One can observe that, in agreement with the intuition, metaorders with shorter durations are more frequent (Fig. \ref{vol atmf T distribution}). This histogram is quite similar to the one of presented for the equity aggressive metaorders in \cite{said2018}.

\paragraph{Daily participation rate distribution}
\label{daily participation rate distribution ATMF volatility metaorders}

\begin{figure}[H]
\centering
\includegraphics[scale=0.80]{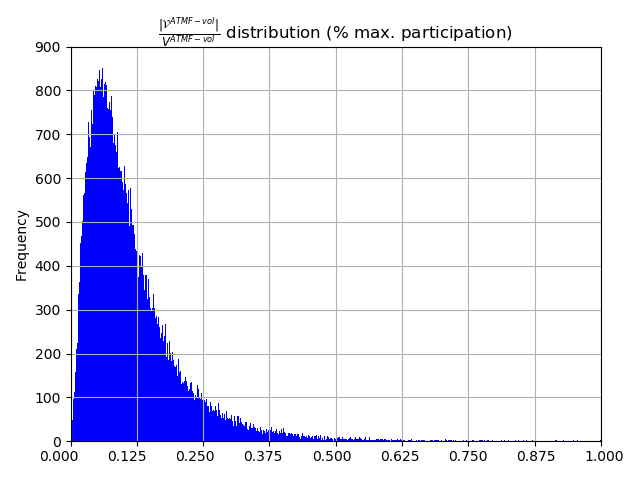}
\captionof{figure}{Daily participation rate distribution of the \textit{at the money forward volatility} metaorders}
\label{vol atmf participation distribution}
\end{figure}

One can note in Figure \ref{vol atmf participation distribution} that metaorders with very high participation rates are evenly represented.

\subsubsection{The ATMF skew metaorders}
\label{the ATMF skew metaorders}

In this section we focus on metaorders whose \textit{at the money forward skew} parameter is the parameter of interest in the implied volatility model.

\paragraph{Data}
\label{data ATMF skew metaorders}

\begin{itemize}
    \item Study period : \textbf{1st July 2016 -- 30th June 2018}
    \item Order types : \textbf{Aggressive Limit Orders}
    \item Parameter : $\theta \equiv$ \textbf{at the money forward skew}
    \item Filters : \textbf{metaorders $\omega \in \Omega^{\theta}$}
    \item Number of orders : \textbf{1,304,714}
    \item Number of metaorders : \textbf{174,091}
\end{itemize}

\paragraph{Duration distribution}
\label{duration distribution ATMF skew metaorders}

\begin{figure}[H]
\centering
\includegraphics[scale=0.80]{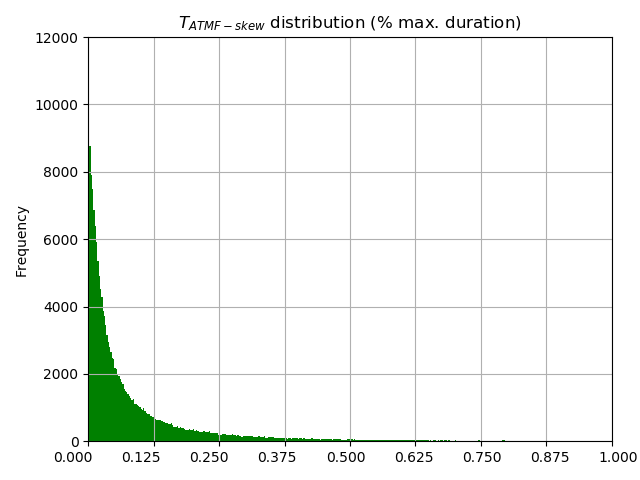}
\captionof{figure}{Duration distribution of the \textit{at the money forward skew} metaorders}
\label{ATMF skew T distribution}
\end{figure}

The distribution of the durations of the \textit{at the money forward skew} metaorders (Fig. \ref{ATMF skew T distribution}) is quite similar to the distribution observed in Figure \ref{vol atmf T distribution} for the \textit{at the money forward volatility} metaorders. One more time we notice that metaorders with shorter durations are more frequent.

\paragraph{Daily participation rate distribution}

\begin{figure}[H]
\centering
\includegraphics[scale=0.80]{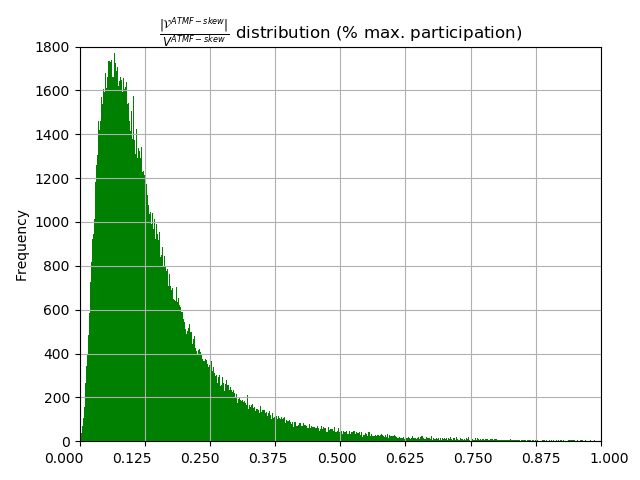}
\captionof{figure}{Daily participation rate distribution of the \textit{at the money forward skew} metaorders}
\label{ATMF skew participation distribution}
\end{figure}

One can note in Figure \ref{ATMF skew participation distribution} that metaorders with very high participation rates are evenly represented. As already observed in Sec. \ref{daily participation rate distribution ATMF volatility metaorders} there is a maximum frequency for metaorders around a certain value closed to the maximum observed in Figure \ref{vol atmf participation distribution}.

\subsubsection{Comparisons of Volatility and Skew metaorders distributions}
\label{comparisons of volatility and skew metaorders distributions}

In Figures \ref{vol atmf T distribution}, \ref{vol atmf participation distribution}, \ref{ATMF skew T distribution} and \ref{ATMF skew participation distribution} durations and daily participation rates are disclosed as a percentage of reference maximum values relative to the market and BNP Paribas options desk. We will denote $\left\langle T\right\rangle_{ref}^{ATMF-vol}$ and $\left\langle T\right\rangle_{ref}^{ATMF-skew}$ these reference durations and $\left\langle\displaystyle\frac{|\mathcal{V}|}{V}\right\rangle_{ref}^{ATMF-vol}$ and $\left\langle\displaystyle\frac{|\mathcal{V}|}{V}\right\rangle_{ref}^{ATMF-skew}$ these reference daily participation rates. As
$$ \left\langle T \right\rangle_{ref}^{ATMF-vol} \approx \left\langle T \right\rangle_{ref}^{ATMF-skew} $$
and
$$ \left\langle\displaystyle\frac{|\mathcal{V}|}{V}\right\rangle_{ref}^{ATMF-vol} \approx \left\langle\displaystyle\frac{|\mathcal{V}|}{V}\right\rangle_{ref}^{ATMF-skew}, $$
we can say that the duration and participation distributions are quite similar for the two types of metaorders. Also the two kinds of metaorders present closed number of orders and metaorders during the same period as disclosed in Sec. \ref{data ATMF volatility metaorders} and in Sec. \ref{data ATMF skew metaorders}, so we can conclude that \textit{at the money forward} volatility and skew metaorders must behave in the same manner. Hence the differences in terms of market impact between the two types of metaorders could be only explained by the way the market reacts to each type of metaorder.

\subsection{Notations}
\label{notations}

\begin{center}
\begin{tabular}{l l}
\toprule[0.15 em]
Notation & \multicolumn{1}{c}{Definition} \\
\midrule[0.1 em]
\centering $\omega$ & \centering A metaorder \tabularnewline
\centering $O(\omega$) & \centering Option of the metaorder $\omega$ \tabularnewline
\centering $d(\omega$) & \centering Execution day of the metaorder $\omega$ \tabularnewline
\centering $t_0(\omega)$ & \centering Start time of the metaorder $\omega$ \tabularnewline
\centering $T(\omega)$ & \centering Duration of the metaorder $\omega$ \tabularnewline
\centering $N(\omega)$ & \centering Length of the metaorder $\omega$ \tabularnewline
\centering $\mathcal{V}^{\theta}(\omega)$ & \centering $\theta-$Sensitivity of the metaorder $\omega$ \tabularnewline
\centering $V^{\theta}(\omega)$ & \centering $\theta-$Sensitivity traded the day $d(\omega)$ on all the options of the universe \tabularnewline
\centering $\epsilon^{\theta}(\omega)$ & \centering Sign of $\mathcal{V}^{\theta}(\omega)$ \tabularnewline
\centering $\mathcal{O}(\omega)$ & \centering Price of $O(\omega)$ \tabularnewline
\centering $\Omega^{\theta}$ & \centering Set of all the $\theta-$metaorders identified by the algorithm \tabularnewline
\centering $\Omega_{n^*}^{\theta} \subset \Omega $ & \centering Subset of the $\theta-$metaorders with $N \geq n{*}$ \tabularnewline
\bottomrule[0.15 em]
\end{tabular}
\captionof{table}{\textit{Notations and definitions}}
\label{tab2}
\end{center}

\begin{rem}
As we only consider metaorders that have at least 5 executed transactions, $\Omega^{\theta} = \Omega_5^{\theta}$.
\end{rem}

\section{Market Impact Dynamics}
\label{market impact dynamics}

\subsection{Market impact curves}
\label{market impact curves}

The main results related to market impact dynamics are now given, namely, the \textit{market impact curves} for the \textit{at the money forward volatility} and the \textit{at the money forward skew} metaorders. In order to plot the market impact dynamics, a similar \textit{bucketing} method as the one presented in \cite{said2018} is used: Let $x$, $y$ being two arrays of data and consider for example that one wants to plot $y$ as a function $x$. First one starts by ordering the couple of values $(x_i, y_i)$ according to the values of $x$ and then divides the \textbf{sorted (by $x$) distribution} $(x,y)_{sorted}$ into $N_{bucket}$ equally-sized buckets. This procedure yields $N_{bucket}$ subsets of the distribution $(x,y)_{sorted}$, $(x_i, y_i)_{i \in I_1}, (x_i, y_i)_{i \in I_2},..., (x_i, y_i)_{i \in I_{N_{bucket}}}$, and for each bucket $I_k$ the \textit{means values} $(\overline{x}_k, \overline{y}_{k})$ is computed. The last step of this bucketing method is to plot the points $(\overline{x}_1, \overline{y}_1), (\overline{x}_2, \overline{y}_2), ...,(\overline{x}_{N_{bucket}}, \overline{y}_{N_{bucket}})$.

To study the dynamics of the market impact, one plots $(\epsilon(\omega)\mathcal{I}_t(\omega))_{\omega \in \Omega, t_0(\omega) \leq\, t \,\leq t_0(\omega) + 2T(\omega)}$. The first sub-interval $t_0(\omega) \leq\, t \,\leq t_0(\omega) + T(\omega)$ corresponds to the execution of the metaorder, whereas the second $t_0(\omega) + T(\omega)  \leq\, t \,\leq t_0(\omega) + 2T(\omega)$ corresponds to the relaxation. The study of relaxation presents a degree of arbitrariness, since a choice has to be made as to the elapsed time after the metaorder is completed. For the sake of homogeneity, the relaxation is measured over the same duration as the execution. This choice seems to be a good compromise to cope with two antagonistic requirements, one being to minimize this elapsed time because of the diffusive nature of prices -- which affect also the parameters --, the other being to maximize it so as to make sure that the relaxation is achieved.

In order to perform an extensive statistical analysis involving metaorders of varying lengths in physical and $\theta-$sensitivity time, a rescaling in time is necessary, see e.g. \cite{bacry2015market} and \cite{said2018}. With this convention, all orders are executed on the time interval $[0,1]$ and parameter relaxation occurs in the time interval  $[1,2]$. For each metaorder $\omega$, one considers $[0,1]$ instead of $[t_0(\omega), t_0(\omega) + T(\omega)]$ $\left( [0,1] = \displaystyle\frac{[t_0(\omega), t_0(\omega) + T(\omega)] - t_0(\omega) }{T(\omega)} \right)$ for the \textit{execution part} of $\omega$ and $[1,2]$ instead of $[t_0(\omega) + T(\omega), t_0(\omega) + 2T(\omega)]$ for the \textit{relaxation part} of $\omega$, and then averages using the \textit{bucketing method} previously described on the time-rescaled $\theta-$sensitivity quantities. In fact the $\theta-$sensitivity time here plays the same role as the volume time in \cite{said2018}. Indeed, traders in the options markets buy and sell more \textit{sensitivities} than the products themselves in order to reduce the expositions of their portfolio i.e. they think and plan not in terms of options (e.g. calls, puts), but rather in terms of sensitivities of their portfolio.

The time variable $t \in [0,1]$ in the Figures of Sec. \ref{market impact dynamics} is actually the \textit{$\theta-$sensitivity time}, i.e., the ratio between the $\theta-$sensitivity of the metaorder already executed at the time of the observation and the total $\theta-$sensitivity of the metaorder - of course, at the end of the execution part this quantity is always equal to 1.

The market impacts disclosed in Figures \ref{vol atmf dynamics N_Vol 5}, \ref{vol atmf dynamics N_Vol 10} and \ref{vol atmf dynamics N_Vol 15} in Sec. \ref{the ATMF volatility market impact dynamics} (resp.  Figures \ref{ATMF skew dynamics N_Smile 5}, \ref{ATMF skew dynamics N_Smile 10} and \ref{ATMF skew dynamics N_Smile 15} in Sec. \ref{the ATMF skew market impact dynamics}) are given in percentage of the same reference value making possible comparisons between them.

\subsection{The ATMF volatility market impact dynamics}
\label{the ATMF volatility market impact dynamics}

The blue points correspond to $\theta-$values during execution, $\theta$ being the \textit{at the money forward volatility} parameter, and the red points correspond to $\theta-$values observed at identical times starting from the end of the metaorder.

\begin{figure}[H]
\centering
\includegraphics[scale=0.80]{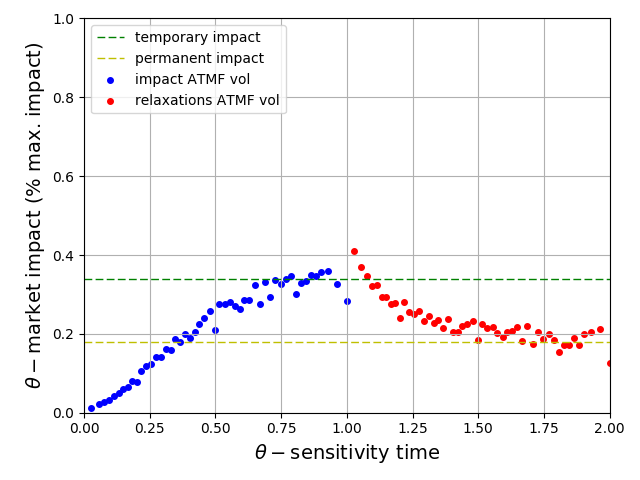}
\captionof{figure}{Market impact dynamics in the case of the \textit{at the money forward volatility} metaorders ($\theta \equiv$ ATMF volatility, set : $\Omega$, 1,026,197 orders, 149,441 metaorders, temporary impact: 0.34, permanent impact: 0.17)}
\label{vol atmf dynamics N_Vol 5}
\end{figure}

\begin{figure}[H]
\centering
\includegraphics[scale=0.80]{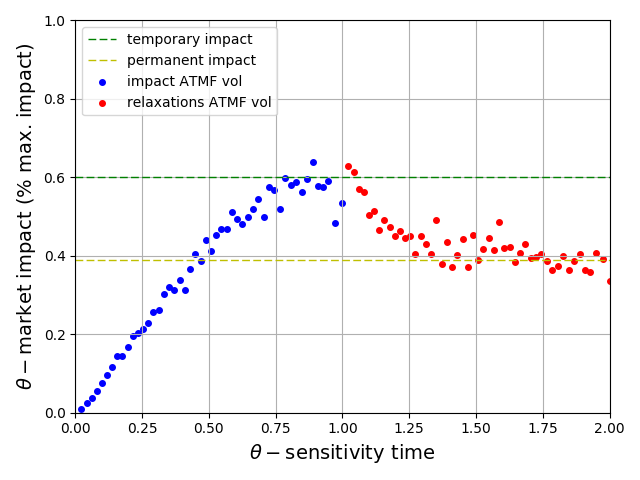}
\captionof{figure}{Market impact dynamics in the case of the \textit{at the money forward volatility} metaorders ($\theta \equiv$ ATMF volatility, set : $\Omega_{10}$, 215,274 orders, 17,286 metaorders, temporary impact: 0.60, permanent impact: 0.39)}
\label{vol atmf dynamics N_Vol 10}
\end{figure}

\begin{figure}[H]
\centering
\includegraphics[scale=0.80]{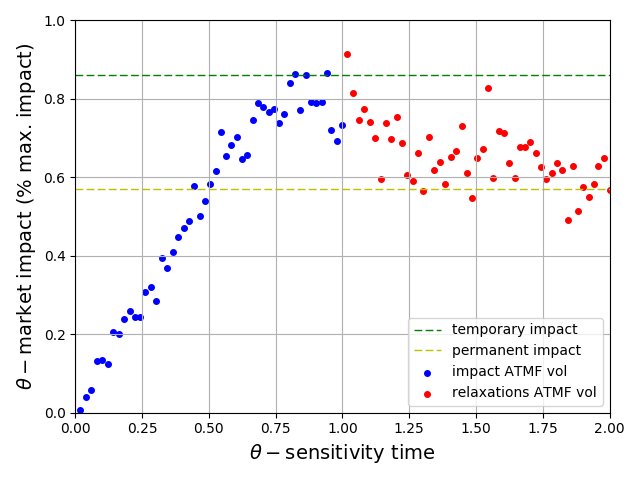}
\captionof{figure}{Market impact dynamics in the case of the \textit{at the money forward volatility} metaorders ($\theta \equiv$ ATMF volatility, set : $\Omega_{15}$, 54,203 orders, 2,958 metaorders, temporary impact: 0.86, permanent impact: 0.57)}
\label{vol atmf dynamics N_Vol 15}
\end{figure}

The analysis clearly yields an increasing, concave market impact curve. However, on can observe that the curve has a linear behavior at the beginning and becomes more concave towards the end. This is explained in particular by the fact that the duration of the metaorders is quite short. The decay observed in the last points (in $t=1.0$ and $t=2.0$) of the curve is an artifact, already discussed in \cite{said2018}, inducing a bias towards the end of the curve. It can be explained by the larger number of metaorders of smaller lengths and with lower impact. Also note that on the three figures \ref{vol atmf dynamics N_Vol 5}, \ref{vol atmf dynamics N_Vol 10} and \ref{vol atmf dynamics N_Vol 15}, the larger the metaorders, the higher the impacts: $0.34$, $0.60$ and then $0.86$ for the temporary market impact.

Figures \ref{vol atmf dynamics N_Vol 5}, \ref{vol atmf dynamics N_Vol 10} and \ref{vol atmf dynamics N_Vol 15} clearly exhibit the concave shape of market impact during the execution part, followed by a convex and decreasing relaxation. Simply by eyeballing Figures \ref{vol atmf dynamics N_Vol 5}, \ref{vol atmf dynamics N_Vol 10} and \ref{vol atmf dynamics N_Vol 15}, one can safely assume that relaxation is complete and stable at a level around respectively $0.17$, $0.39$ and $0.57$. However, on Figure \ref{vol atmf dynamics N_Vol 15}, relaxation does not seem to be quite smooth. This behaviour for those larger metaorders is essentially  due to the fact that the set $\Omega_{15}$ contains much less metaorders.

A conclusion to this section is that the concave shape of the temporary impact and the convex relaxation curve concerning the \textit{at the money forward volatility} metaorders are in line with the empirical results observed on the equity markets and higlighted in \cite{bacry2015market}, \cite{bershova2013non} and \cite{said2018}. Also, and more interestingly, the market impact and relaxation curves confirm the theoretical findings of \cite{farmer2013efficiency} that the impact should be concave and increasing, and that the final impact after the execution is performed should relax to about two-thirds of the peak impact. Indeed while on the Figure \ref{vol atmf dynamics N_Vol 5} the ratio between the \textit{permanent market impact} and the \textit{temporary market impact} seems to be much closer to $1/2$, one can observe in Figures \ref{vol atmf dynamics N_Vol 10} and \ref{vol atmf dynamics N_Vol 15} -- which correspond to larger metaorders and therefore more significant --, how this ratio gets close to $2/3$.

\subsection{The ATMF skew market impact dynamics}
\label{the ATMF skew market impact dynamics}

The main results related to skew market impact dynamics are now given, namely, the \textit{market impact curves} for the \textit{at the money forward skew} metaorders. To this purpose we use the same \textit{bucketting method} introduced in Sec. \ref{market impact curves}.

\begin{figure}[H]
\centering
\includegraphics[scale=0.80]{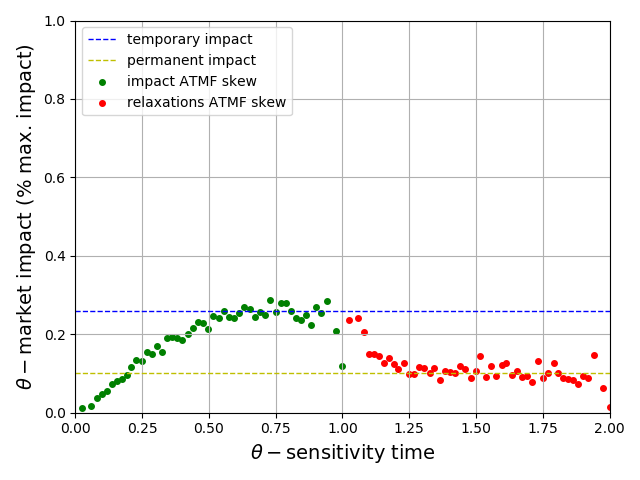}
\captionof{figure}{Market impact dynamics in the case of the \textit{at the money forward skew} metaorders ($\theta \equiv$ ATMF skew, set : $\Omega$, 1,304,714 orders, 174,091 metaorders, temporary impact: 0.26, permanent impact: 0.10)}
\label{ATMF skew dynamics N_Smile 5}
\end{figure}

The green points correspond to $\theta-$values during execution, $\theta$ being the \textit{at the money forward skew} parameter, and the red points correspond to $\theta-$values observed at identical times starting from the end of the metaorder.

\begin{figure}[H]
\centering
\includegraphics[scale=0.80]{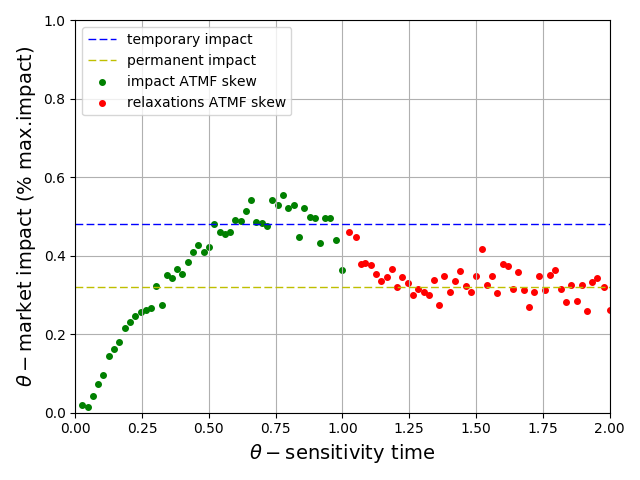}
\captionof{figure}{Market impact dynamics in the case of the \textit{at the money forward skew} metaorders ($\theta \equiv$ ATMF skew, set : $\Omega_{10}$, 405,918 orders, 30,932 metaorders, temporary impact: 0.48, permanent impact: 0.32)}
\label{ATMF skew dynamics N_Smile 10}
\end{figure}

\begin{figure}[H]
\centering
\includegraphics[scale=0.80]{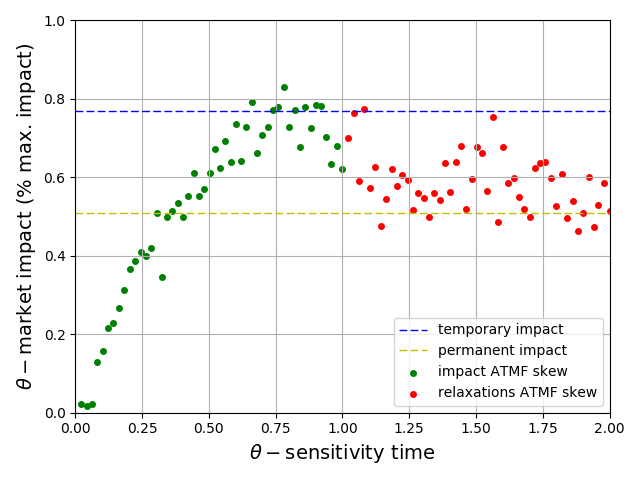}
\captionof{figure}{Market impact dynamics in the case of the \textit{at the money forward volatility} metaorders ($\theta \equiv$ ATMF skew, set : $\Omega_{15}$, 136,410 orders, 7,260 metaorders, temporary impact: 0.77, permanent impact: 0.51)}
\label{ATMF skew dynamics N_Smile 15}
\end{figure}

The results show an increasing, concave market impact curve, with a linear behavior at the beginning. The curve becomes more concave towards the end. As mentioned in Sec. \ref{the ATMF volatility market impact dynamics}, this is due to the fact that the durations of the metaorders are quite short. The decay observed in the last points (in $t=1.0$ and $t=2.0$) is the same effect observed in the \textit{at the money forward volatility} metaorders. Also note that on the three figures \ref{ATMF skew dynamics N_Smile 5}, \ref{ATMF skew dynamics N_Smile 10} and \ref{ATMF skew dynamics N_Smile 15}, the larger the metaorders, the higher the impacts: $0.26$, $0.48$ and then $0.77$ for the temporary market impact.

Figures \ref{ATMF skew dynamics N_Smile 5}, \ref{ATMF skew dynamics N_Smile 10} and \ref{ATMF skew dynamics N_Smile 15} clearly exhibit the concave shape of market impact during the execution part, followed by a convex and decreasing relaxation. Simply by eyeballing Figures \ref{ATMF skew dynamics N_Smile 5}, \ref{ATMF skew dynamics N_Smile 10} and \ref{ATMF skew dynamics N_Smile 15}, one can safely assume that relaxation is complete and stable at a level around respectively $0.10$, $0.32$ and $0.51$. However, on Figure \ref{ATMF skew dynamics N_Smile 15}, relaxation does not seem to be quite smooth. This behaviour for those larger metaorders is essentially  due to the fact that the set $\Omega_{15}$ contains much less metaorders.

A conclusion to this section is that the concave shape of the temporary impact and the convex relaxation curve concerning the \textit{at the money forward skew} metaorders are in line with the empirical results observed on the \textit{at the money forward volatility} metaorders. Also, and more interestingly, the market impact and relaxation curves confirm the theoretical findings of \cite{farmer2013efficiency} that the impact should be concave and increasing, and that the final impact after the execution is performed should relax to about two-thirds of the peak impact. Indeed while on the Figure \ref{ATMF skew dynamics N_Smile 5} the ratio between the \textit{permanent market impact} and the \textit{temporary market impact} seems to be much closer to $0.4$, one can observe in Figures \ref{vol atmf dynamics N_Vol 10} and \ref{vol atmf dynamics N_Vol 15} -- which correspond to larger metaorders and therefore more significant --, how this ratio gets close to $2/3$.

\section{Square-Root Law}
\label{square root law}

\subsection{Volatility and Skew metaorders square-root laws}
\label{volatility and skew square-root laws}

The results presented in this section are certainly the most important of the article. They confirm the consistency of the \textit{Square-Root Law} already observed in the equity market \cite{almgren2005direct}, \cite{bershova2013non}, \cite{gomes2015market}, \cite{mastromatteo2014agent}, \\\cite{moro2009market} and \cite{toth2011anomalous}, the bitcoin market \cite{donier2015million} and more recently in a short note \cite{toth2016square} the authors highlighted that the \textit{Square-Root Law} also holds for option markets according to their definition of the \textit{implied volatility} metaorders which is more global. The method presented here which relies on the definition \ref{defMetaorderOption} tends to be more local focusing on the local deformations of the implied volatility surface through the variations of the parameters of the model which are in fact the projections of those local deformations.

The \textit{Square-Root Law} is the fact that the impact curve should not depend on the duration of the metaorder. Indeed, almost all studies now agree on the fact that the impact is more or less close to be proportional to the square root of the volume executed. Considering options market, which plays the role of an executed volume is in fact the $\theta-$sensitivity executed regarding to a $\theta-$metaorder.  
However, the so-called \textit{Square-Root Law} states much more than that. It basically claims that the market impact does not depend on the metaorder duration.

\begin{figure}[H]
\centering
\includegraphics[scale=0.80]{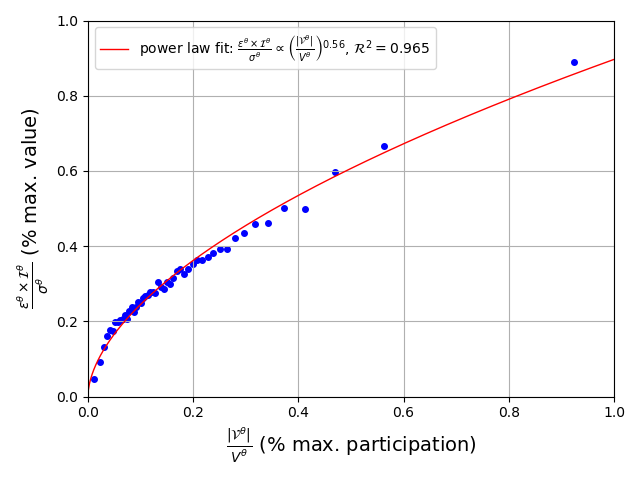}
\captionof{figure}{The square-root law in the case of the \textit{at the money forward volatility} metaorders, power law fit: $y \propto x^{0.56}$, $\mathcal{R}^2 = 0.965$.}
\label{square root law vol atmf curve}
\end{figure}

\begin{figure}[H]
\centering
\includegraphics[scale=0.80]{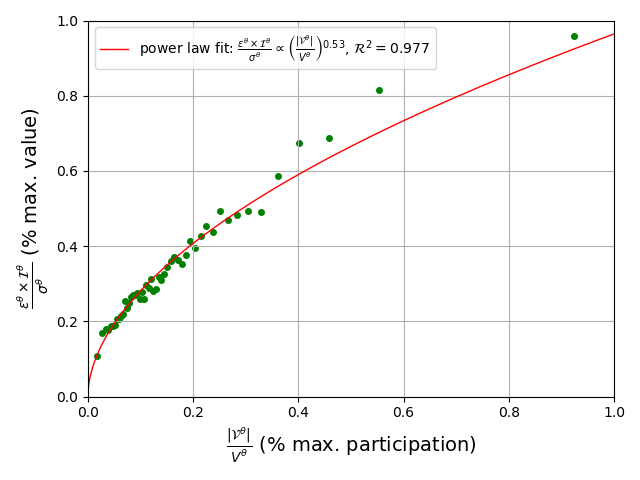}
\captionof{figure}{The square-root law in the case of the \textit{at the money forward skew} metaorders, power law fit: $y \propto x^{0.53}$, $\mathcal{R}^2 = 0.977$.}
\label{square root law ATMF skew curve}
\end{figure}

In what follows one plots the $\theta-$market impact normalized by a volatility factor $\epsilon^{\theta}\times\displaystyle\frac{\mathcal{I}^{\theta}}{\sigma^{\theta}}$ as a function of the $\theta-$daily participation rate
$\displaystyle\frac{\left|\mathcal{V}^{\theta}\right|}{V^{\theta}}$ for both the \textit{at the money forward volatility} (Fig. \ref{square root law vol atmf curve}) and the \textit{at the money forward skew} (Fig. \ref{square root law ATMF skew curve}) metaorders, $\sigma^{\theta}$ being the daily standard deviation of the parameter $\theta$ and $\mathcal{I}_t^{\theta} = \theta_t - \theta_{t_0}$ as defined in Equation (\ref{variation proxy}). One observes that a power-law fit gives exponents for the $\theta-$daily participation rate close to $0.5$ ($\approx 0.56$ for the \textit{at the money forward volatility} metaorders and $\approx 0.53$ for the \textit{at the money forward skew} metaorders).
In both cases, the analysis shows that our options metaorders, present a market impact following a theoretical curve of the form $\sigma \sqrt{\mathcal{R}}$ with $\sigma$ a volatility factor and $\mathcal{R}$ a participation rate factor. Those findings support the idea for a universal underlying mechanism in market microstructure.

\subsection{Comparisons of Volatility and Skew metaorders square-root laws}
\label{comparisons of volatility and skew metaorders square-root laws}

As already done in Sec. \ref{comparisons of volatility and skew metaorders distributions} we are going now to make some comparisons between the square-root laws presented in Sec. \ref{volatility and skew square-root laws}. In Figures \ref{square root law vol atmf curve} and \ref{square root law ATMF skew curve} normalized market impacts and daily participation rates are disclosed in percentage of reference maximum values. We will denote respectively $\left\langle\displaystyle\frac{\epsilon \times \mathcal{I}}{\sigma}\right\rangle_{ref}^{ATMF-vol}$, $\left\langle\displaystyle\frac{\epsilon \times \mathcal{I}}{\sigma}\right\rangle_{ref}^{ATMF-skew}$, $\left\langle\displaystyle\frac{|\mathcal{V}|}{V}\right\rangle_{ref}^{ATMF-vol}$ and $\left\langle\displaystyle\frac{|\mathcal{V}|}{V}\right\rangle_{ref}^{ATMF-skew}$ these reference values. The following relations
$$ \left\langle\displaystyle\frac{\epsilon \times \mathcal{I}}{\sigma}\right\rangle_{ref}^{ATMF-vol} \approx 2 \times \left\langle\displaystyle\frac{\epsilon \times \mathcal{I}}{\sigma}\right\rangle_{ref}^{ATMF-skew} $$
and
$$ \left\langle\displaystyle\frac{|\mathcal{V}|}{V}\right\rangle_{ref}^{ATMF-vol} \approx 0.7 \times \left\langle\displaystyle\frac{|\mathcal{V}|}{V}\right\rangle_{ref}^{ATMF-skew}, $$
show in comparison to skew metaorders, that lower participation rates is needed to obtain larger normalized impacts in volatility metaorders. Hence altough volatility and skew metaorders present similarities as underlined in Sec. \ref{comparisons of volatility and skew metaorders distributions} the market reacts much more easily to metaorders of volatility. This can be explained by the fact that \textit{trading volatility} is more common than \textit{trading skew} in options market.

\section{Fair Pricing}
\label{fair pricing}

\begin{figure}[H]
\centering
\includegraphics[scale=0.80]{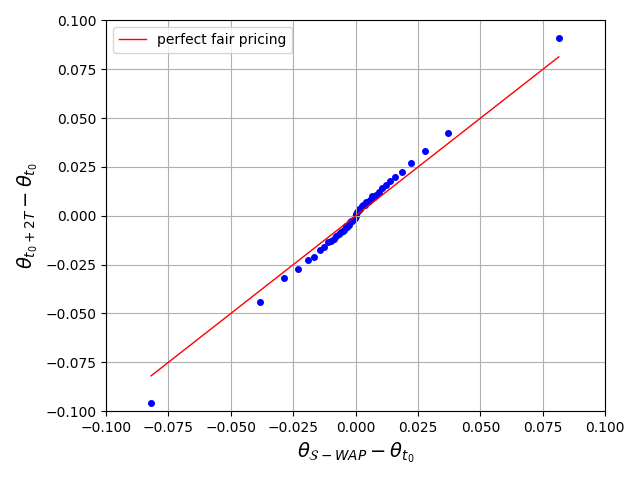}
\captionof{figure}{Fair pricing of the \textit{at the money forward volatility} parameter in the case of the \textit{at the money forward volatility} metaorders}
\label{fair pricing vol atmf}
\end{figure}

\begin{figure}[H]
\centering
\includegraphics[scale=0.80]{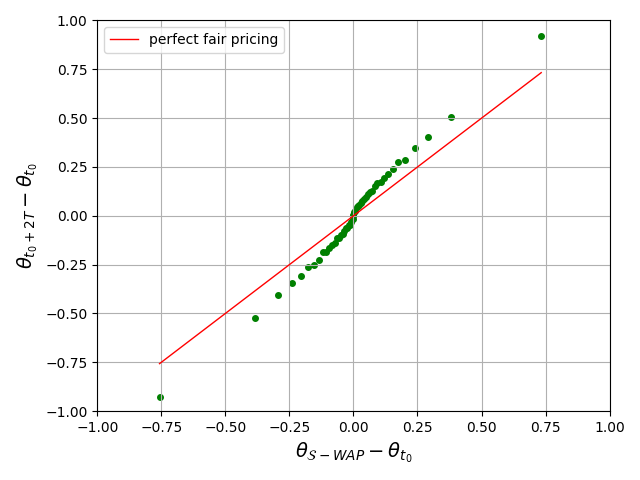}
\captionof{figure}{Fair pricing of the \textit{at the money forward skew} parameter in the case of the \textit{at the money forward skew} metaorders}
\label{fair pricing ATMF skew}
\end{figure}

In this section we deal with the fair pricing condition of our options metaorders. First of all let us define the $\mathcal{S}$-WAP (sensitivity weighted average parameter) of a metaorder $\omega$ as the quantity defined by 
$$ \theta_{\mathcal{S}-WAP}(\omega) = \displaystyle\frac{\displaystyle\sum_{i=0}^{N(\omega)-1}{\mathcal{S}_i^{\theta}(\omega) \theta_{t_i(\omega)}(\omega)}}{\mathcal{V}^{\theta}(\omega)} $$
where $t_0(\omega), ..., t_{N(\omega)-1}(\omega)$ represent the times of the transactions of the metaorder $\omega$ and $\mathcal{V}^{\theta}(\omega) = \displaystyle\sum_{i=0}^{N(\omega)-1}{\mathcal{S}_i^{\theta}(\omega)}$. Hence we want to compare $\theta_{\mathcal{S}-WAP} - \theta_{t_0}$ with $\theta_{t_0 + 2T} - \theta_{t_0}$ (Fig. \ref{fair pricing vol atmf} and \ref{fair pricing ATMF skew}). The red line represents the perfect fair pricing condition as it corresponds to $\theta_{\mathcal{S}-WAP} - \theta_{t_0} = \theta_{t_0 + 2T} - \theta_{t_0} $.

It appears from Figures \ref{fair pricing vol atmf} and \ref{fair pricing ATMF skew} that the fair pricing condition can reasonably be assumed to hold. This is in line with what has already been observed in the equity market and mentioned in \cite{said2018}. One observes also that the greater the absolute $\theta-$variations, the more one moves away from the perfect fair pricing condition. In agreement with the intuition high variations are generally associated to longer and larger metaorders that are therefore more affected by the diffusive nature of the prices.

One can note that the fair pricing condition is more effective for the \textit{at the money forward volatility} metaorders. One reason for this is that trading \textit{volatility} is more common than trading \textit{skew} on the options market as already mentionned previously.

\begin{figure}[H]
\centering
\includegraphics[scale=0.80]{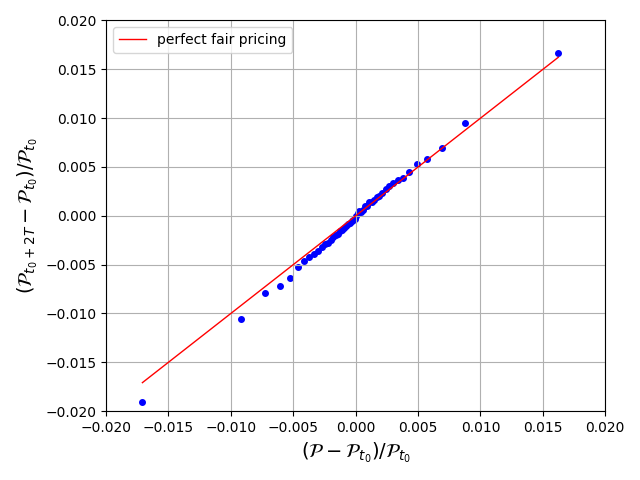}
\captionof{figure}{Fair pricing of the portfolio value in the case of the \textit{at the money forward volatility} metaorders}
\label{fair pricing portfolio vol atmf}
\end{figure}

\begin{figure}[H]
\centering
\includegraphics[scale=0.80]{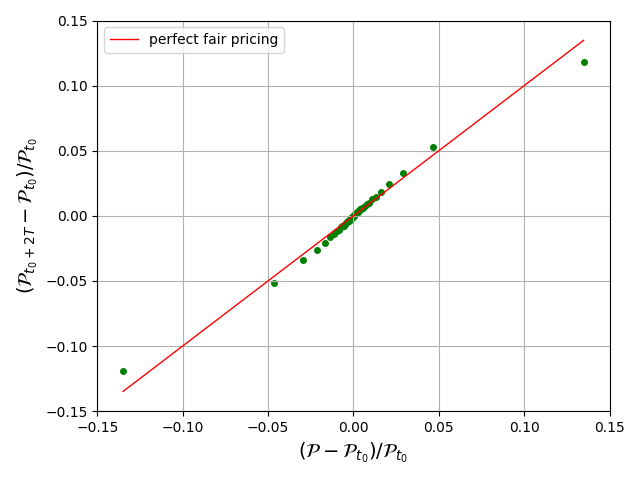}
\captionof{figure}{Fair pricing of the portfolio value in the case of the \textit{at the money forward skew} metaorders}
\label{fair pricing portfolio ATMF skew}
\end{figure}

For now, we have considered the fair pricing condition by studying the variations of the parameters of the metaorders in question. However, it is more relevant to examine the fair pricing condition by considering the portfolio generated during a metaorder. Let us define the portfolio value of a metaorder $\omega$ as the quantity defined by 
$$ \mathcal{P}(\omega) = \displaystyle\sum_{i=0}^{N(\omega)-1}{|Q_i(\omega)| \mathcal{O}_{t_i(\omega)}(\omega)} $$
where $t_0(\omega), ..., t_{N(\omega)-1}(\omega)$ represent the instants, $|Q_0(\omega)|, ..., |Q_{N(\omega)-1}(\omega)|$ the quantity (positive) and  $\mathcal{O}_0(\omega), ..., \mathcal{O}_{N(\omega)-1}(\omega)$ the prices of the transactions of the metaorder $\omega$. Hence we want to compare $\displaystyle\frac{\mathcal{P} - \mathcal{P}_{t_0}}{\mathcal{P}_{t_0}}$ with $\displaystyle\frac{\mathcal{P}_{t_0 + 2T} - \mathcal{P}_{t_0}}{\mathcal{P}_{t_0}}$ (Fig. \ref{fair pricing portfolio vol atmf} and \ref{fair pricing portfolio ATMF skew}) where $\mathcal{P}_{t_0}$ and $\mathcal{P}_{t_0 + 2T}$ are respectively the prices of the same portfolio at $t_0$ and $t_0 + 2T$. The red line represents the perfect fair pricing condition as it corresponds to $\displaystyle\frac{\mathcal{P} - \mathcal{P}_{t_0}}{\mathcal{P}_{t_0}} = \displaystyle\frac{\mathcal{P}_{t_0 + 2T} - \mathcal{P}_{t_0}}{\mathcal{P}_{t_0}} $.

To conclude this section, we observe that the fair pricing seems to be also hold on the options market. This confirms the fair pricing hypothesis introduced in \cite{farmer2013efficiency} as a universal mechanism concerning the metaorders and their interaction with the price formation process. 

\section{Conclusion}
\label{conclusion}

This work is an empirical study of a large set of metaorders in one of the main Asian index options market. A new algorithmic definition of an \textit{options metaorder} has been proposed. Our study contains two distinct groups of metaorders using aggressive limit orders: a set of \textit{at the money forward volatility} metaorders, and a database of \textit{at the money forward skew} metaorders. The statistical results based on this definition show a pretty good agreement with some observations already highlighted in the stock markets: \textit{Square-Root Law}, \textit{Fair Pricing} and \textit{Market Impact Dynamics}. In both cases, the analysis shows that the temporary impact is increasing and concave, with a convex decreasing relaxation phase. More precisely, the price reversion after the completion of a trade yields a permanent impact such that its ratio to the maximum impact observed at the last fill is roughly two-third, as predicted in the paper of \cite{farmer2013efficiency} and already highlighted empirically on equity markets.

\newpage
\bibliographystyle{apalike}
\bibliography{bibliography}

\end{document}